\documentclass[]{revtex4-1}
\usepackage{docs}%
\usepackage{bm}%
\usepackage[colorlinks=true,linkcolor=blue]{hyperref}%

\usepackage{graphicx}
\usepackage[english]{babel}
\usepackage[export]{adjustbox}
\usepackage{booktabs}
\usepackage{textcomp}
\usepackage{amsmath}

\usepackage{xcolor}
\usepackage{ulem}

\begin{document}

\title{LION : \Large {L}aser {I}nterferometer {O}n the moo{N}}

\author{Pau Amaro-Seoane}
\affiliation{Universitat Polit{\`e}cnica de Val{\`e}ncia, IGIC, Spain}
\affiliation{DESY, Zeuthen, Germany}
\affiliation{Kavli Institute for Astronomy and Astrophysics, China}
\affiliation{Institute of Applied Mathematics, Academy of Mathematics and Systems Science, China}
\affiliation{Zentrum f{\"u}r Astronomie und Astrophysik, TU Berlin, Germany}

\author{Lea Bischof}
\affiliation{ Max Planck Institute for Gravitational Physics (Albert Einstein Institute)}
\affiliation{ Institute for Gravitational Physics of the Leibniz Universit\"at Hannover}
 
\author{Jonathan J. Carter}
\affiliation{ Max Planck Institute for Gravitational Physics (Albert Einstein Institute)}
\affiliation{ Institute for Gravitational Physics of the Leibniz Universit\"at Hannover}

\author{Marie-Sophie Hartig}
\affiliation{ Max Planck Institute for Gravitational Physics (Albert Einstein Institute)}
\affiliation{ Institute for Gravitational Physics of the Leibniz Universit\"at Hannover}

\author{Dennis Wilken}
\affiliation{ Max Planck Institute for Gravitational Physics (Albert Einstein Institute)}
\affiliation{ Institute for Gravitational Physics of the Leibniz Universit\"at Hannover}

\def\apjl{Astrophys. J. Lett.}

\date{\today ~ - All authors contributed equally to this work.}

\begin{abstract}

Gravitational wave astronomy has now left its infancy and has become an
important tool for probing the most violent phenomena in our universe. The
LIGO/Virgo-KAGRA collaboration operates ground based detectors which cover the
frequency band from 10\,Hz to the kHz regime. Meanwhile, the pulsar timing array and
{the} soon to launch LISA mission will cover frequencies below 0.1\,Hz, leaving
a gap in detectable gravitational wave frequencies.  Here we show how a Laser Interferometer On the mooN (LION) gravitational wave detector
would be sensitive to frequencies from sub\,Hz to
kHz. 
We find that the sensitivity curve is such that LION can measure compact
binaries with masses between $10$ and $100\,M_{\odot}$ at cosmological
distances, with redshifts as high as $z=100$ and beyond, depending on the spin
and the mass ratio of the binaries. LION can detect binaries of compact objects with
higher-masses, with very large signal-to-noise ratios, help
us to understand how supermassive black holes got their colossal masses on the
cosmological landscape, and it can observe in
detail intermediate-mass ratio inspirals at distances as large as at least
$100\,\textrm{Gpc}$. Compact binaries that never reach the LIGO/Virgo
sensitivity band can spend significant amounts of time in the LION band, while
sources present in the LISA band can be picked up by the detector and observed
until their final merger. Since LION covers the deci-Hertz regime with such
large signal-to-noise ratios, it truly achieves the dream of multi messenger
astronomy. 

\end{abstract}

\maketitle

\section{Introduction}
{ The LIGO-Virgo-KAGRA (LVK) collaboration \cite{aligo,caron1997virgo,aso2013interferometer} now forms a network of gravitational wave observatories covering the Earth. 
Third generation ground based detectors such as the Einstein Telescope (ET)~\cite{Punturo_2010} and Cosmic Explorer (CE)~\cite{PhysRevD.91.082001,reitze2019cosmic} are currently in the planning stage of development. 
These detectors are planned to push the lower frequency bound of Earth based gravitational wave detectors to their fundamental limits.}
\par
{Newtonian noise, or gravitational gradient noise, forms a lower frequency bound of Earth based detection at about 1$\,$Hz \cite{Beker_2012}.
Meanwhile, with a proposed launch of 2034, the ESA's LISA mission \cite{Amaro-SeoaneEtAl2017} will cover the \textmu Hz to sub deci-Hz  regime of the gravitational wave spectrum. The upper sensitivity bound of space based detectors is set by the interferometer arm length \cite{PhysRevD.96.084004,Schilling_1997}}. 
{These barriers create a gap of measurable frequencies which requires a different style of mission to fill. In this gap many interesting stellar phenomena go unrecorded and their immeasurable impact on the field of gravitational wave astronomy is lost \cite{Manchester_2005}.
Therefore, there is great interest in any gravitational wave detector capable of bridging this gap.
Several missions, at various stages of development, are attempting to do this, but none is yet close to being launched \cite{kawamura2006japanese,phinney2003big}}.
In this work we show a lunar gravitational wave observatory would be able to observe frequencies down to 0.7 Hz while simultaneously overlapping with the sensitive frequency range of Earth based detectors. 
\par

{ During the 90's Wilson and La Fave published two papers on the idea of lunar interferometers, but as the field was still in its infancy, a lot of benefits and challenges went overlooked \cite{wilson93, wilson94}.}
A few other groups have suggested using seismometer arrays on both the Moon and other extraterrestrial solar bodies \cite{PAIK2009167,PhysRevD.90.102001}. 
Coughlin and Harms managed to put limits on the energy contained within the stochastic background across the frequency band 0.1-1\,Hz using seismometers on the Moon from the Apollo missions as a detector. 
A recent proposal has now been submitted to launch many high sensitivity seismometers to the surface of the Moon to push this detection method to its limits \cite{harms2020lunar}. If launched, these seismometers would enable a better understanding of the Moon's seismic activity and so the true limits of seismic noise as discussed later in this paper. 
Aside from that, a paper has been submitted to arxiv recently discussing the concept of a lunar gravitational wave detector \cite{jani2020gravitational}. They focus largely on the astrophysical sources in their sensitivity band. Although their optimistic estimate of performance was excellent at low frequency, it relied upon the use of a freely floating test mass, a technology that has only been proposed and has no further development so far \cite{Friedrich_2014}. 
The conservative estimate they make is similar to the one we reach here, without focusing on the technical details needed to reach that. We also show significant differences in detector design. Still, it is very promising that two independent approaches can reach similar noise estimates. 
\par
In the following we discuss how a ground based long arm interferometer with Fabry-Pérot cavities could be constructed for use as a gravitational wave detector on the surface of the Moon. We will discuss the merits and potential design style of such a lunar observatory.
We address this by constructing a feasible noise budget for a laser observatory based upon current gravitational wave detectors.
We benefit from the lunar surface's seismically quiet environment \cite{Latham455,latham1972passive}, and its lack of atmosphere. Thus the limit on low frequency detection set by the Newtonian noise can be avoided.
By means of this noise budget, we show that the sensitivity gap between the high frequency ground based detectors and low frequency space based detectors can be partially bridged. 
Furthermore, a detector built on the Moon with a sensitivity band also overlapping that of the Earth based network increases the size of the network by a factor of 60, improving the source localisation ability significantly.
\par
Several stellar phenomena are expected in the frequency band exclusively covered by our proposed lunar detector and we discuss the impact their observations will have.
A low frequency detector allows the tracking of merging binaries for much longer times \cite{Maggiore2008}, detecting binaries at much greater redshift \cite{ColpiDotti2011}, detecting heavier binaries \cite{Amaro-SeoaneSantamaria10}, allows for better sky localisation \cite{wen2010geometrical}, and detecting new types of astrophysical sources \cite{Mandel_2018,Amaro-Seoane2018}. 
Long standing questions about the origins of many sources could uniquely be resolved witha a LION detector  \citep{ChenAmaro-Seoane2017}. Furthermore, sub-Hertz detection allows for the probing of gravity in the strong field regime, testing general relativity to its limits \citep{Amaro-SeoaneLRR,Amaro-Seoane2020,Amaro-SeoaneEtAl07}. 
Finding sources at very high redshifts leads directly to a better understanding of the  early universe, where our current understanding is limited by our detection capabilities.\par
{With the renewed interest in lunar missions from several major global powers \cite{mitrofanov2014luna,sibeck2011artemis,xiao2014china,haruyama2012exploration}, now is the ideal time for the gravitational wave community to begin considering what use the Moon can be in aiding our understanding of the cosmos.\par
Section \ref{techdet} starts with the discussion of technical details, including detector parameters, noise contributions and the location and infrastructure. This is followed in Section \ref{sec:NoiseBug} by a summarised noise budget. The science case in Section \ref{sec:Science} discusses the new possibilities that a lunar detector offers, while  Section \ref{sec:costandTime} gives an estimate about the mission cost and timeline.
}

\section{Technical Discussion}
\label{techdet}
The model for our suggested Laser Interferometer On the mooN (LION) is based on the Cosmic Explorer 2 proposal. Our modifications are listed in Table \ref{tab:keyfigs}. These modification arise from the special conditions on the Moon  and design choices made to optimise the frequency band of interest which we discuss below.
 \begin{table}[]
 \setlength{\tabcolsep}{12pt}
     \centering
     \begin{tabular}{l l }\toprule
         \textbf{Parameter} &  \textbf{Value}\\\midrule
         Test Mass (kg) & 1\,267, 1\,267, 698, 707 \\ 
         Suspension length (m) & 3.54, 2.05, 1.66, 2.50 \\ 
         Temperature (K)&  70 \\
         Laser Power (W) &  250 \\
         Wavelength (nm) & 2000\\
         Arm Length (km) &  40 \\
         Seismic & Earth Surface/1000 \\ \bottomrule
        \end{tabular}
     \caption{Overview of the key parameters of LION. All other values correspond to the Cosmic Explorer 2 design. The masses and suspension length are listed from test mass to top mass.}
     \label{tab:keyfigs}
 \end{table}
\par
\subsection{Location}
The choice of the location for LION is predominately determined by the temperature map of the Moon.
A crater near a lunar pole is considered by us to be a suitable position for several reasons; thermal stability, geometry, and shelter from solar radiation and dust \cite{TempCrater,TempMoon1, TempMoon2}.
A promising candidate site is the Shoemaker crater (88.1$^\circ $S 44.9$^\circ $E) at the Moon's south pole.  Hayne et al. \cite{TempCrater} investigated the measurement data of the Lunar Reconnaissance Orbiter (LRO) mission launched in 2009 for three craters with about $50$\,km diameter.
The shoemaker crater shows an average temperature of about $50\mathrm{K}$ with a relatively small variation of minimal/maximal temperature $20-95\mathrm{K}$ compared to other parts of the moon. 
It has a diameter of $50$\,km, introducing one of the limiting factors for the detectors: the arm length.
The nearby Malapert mountain is a nearly perfect site to build solar panels and a communication equipment for data transfer \cite{poweratpoles}.
Additionally, the surface appears to be nearly free of water frost, which could otherwise evaporate onto optics \cite{TempCrater}.
We assume the use of this crater for all future calculations.
\subsection{Infrastructure}
The vacuum on the surface of the Moon is three orders of magnitude better than that in the vacuum tubes in the aLIGO detectors \cite{moonfact,aligo}, and so no infrastructure will be built to house a vacuum. Only the optic support stations are needed to isolate the test optics and allow for their precise alignment controls. 
\par
For the detection of gravitational waves, coincidence data from two independent simultaneous instruments is required.
This is solved in the ET detector and LISA by using a triangular shape with two interferometers in the arms \cite{Punturo_2010,Danzmann1996,McNamara2008}.
Therefore, we propose to build LION in a triangular shape with an arm length of 40\,km, limited by the dimensions of the crater.
\subsection{Background Seismic}
{The seismic noise on the Moon arises from two main sources; meteorite impacts \cite{Apollo1170, Moonquakes80, MeteoriticSeismic09, seismic07} and from the shallow seismic activity of the Moon \cite{nature19}. Many of the sources of seismic activity on Earth do not exist on the Moon \cite{beker2011improving} such as moving cars and trains, winds, and the oceans tides. In particular, the oceans tides cause a large microseismic peak at 0.1-0.2\,Hz which proves troublesome to all detectors seeking low frequency performance; a lunar site completely negates this \cite{matichard2015advanced}. 
Meanwhile, wind and ground conditions on the Earth tilts the detectors, polluting the measurements of inertial sensors with additional noise, which is directly injected into platform controls \cite{lantz2009requirements,Arai_2002}.  The absence, or at the very least substantial reduction, of tilt would allow for a much more aggressive active platform control scheme to reduce motion in the test masses \cite{Carter_2020}. The control noise is currently one of the reasons the aLIGO detectors do not reach their design senstivity at low frequencies.
\par
The only direct seismic measurements on the Moon's surface were performed during the Apollo missions \cite{Latham455, Apollo1170}. They concluded that the background seismic level was significantly below the sensitivity of their measurement devices (0.3\,nm/$\sqrt{\text{Hz}}$ at 1\,Hz), and thus lower than anything on Earth's surface. From this we can safely assume the seismic on the Moon to be smaller by a factor of 1\,000 than Earth's surface seismic motion. However, the Russian operated Luna 27 will be launched in 2024 \cite{Dacey_2019} with a more sensitive seismometer \cite{2011LPI....42.1798M}, possibly measuring an even lower boundary.  }

\subsection{Newtonian Noise}
    Newtonian noise is caused by the movement of mass around the detector causing a change in the local gravitational field \cite{beker2011improving}. These largely come form two sources; the atmosphere and the ground. The lack of atmosphere on the Moon negates the former entirely. 
    Reliable modelling of seismic Newtonian noise for the lunar surface is impossible without accurate measurements of the seismic activity and surface composition. On Earth the noise forms a noise cliff at approximately 1\,Hz. Since the Moon has lower activity \cite{Moonquakes80} and has a lower density on its surface it is assumed that this will occur at a lower frequency. {Newtonian noise is therefore not plotted in the noise budget.} A full study of the gradient noise would still be essential as more knowledge of the Moon's activity is gained.

\subsection{Suspension Systems}
  The limiting low frequency noise considerations are the residual of seismic activity and suspension thermal noise.
 Like in Earth bound detectors, a multistage seismic isolation system would be necessary. This will compose of a preisolation  stage and several suspended masses each acting as a pendulum \cite{cumming2012design, aston2012update}. In order to achieve the required seismic isolation the resonance frequencies of the pendula must be tuned to lower frequency than Cosmic Explorer. The gravity on the Moon's surface is 1.62\,m/s$^2$, approximately a sixth of the gravity on Earth. Although the lower gravity of the Moon offers lower natural frequency by a factor of 0.4, longer suspensions with heavier test masses must still be included. As the gravity is weaker, the masses can be a factor of 6 larger than Cosmic Explorer without additional weight on the suspensions. Making such large test masses and coating them may prove difficult so we assume only a factor of 4 increase can be made. This leads to a test and penultimate mass of 1270\,kg and another 2 stages with about 700\,kg mass. Making the lengths of suspensions on average a factor of 3 longer allows for the whole system to fit in a 11m structure while reducing the noise to a level suitable for sub Hz measurements. Suspension systems are a limiting factor for the low frequency performance of all detectors with fundamental problems of how much weight can be suspended and resonance frequency. The lower gravity on the Moon provides help on both of these fronts. How far this can be developed will form the lower {boundary} of measurable frequencies, but ultimately it is a matter of engineering. 
 \par
\subsection{Thermal Noise}
{Thermal noise is a considerable problem for the low frequency response of any detector. The extremely low temperatures at the Moon's south pole are a good start for achieving the necessary thermal noise suppression. 
 However, a low mechanical loss is also a necessity for a low thermal noise \cite{saulson1994fundamentals}.
 The bulk material of the test masses at both LIGO and Virgo are fused silica, which due to a mechanical loss peak, has poor mechanical loss at cryogenic temperatures \cite{schroeter2007mechanical}. Research for alternative test mass materials is ongoing for use in the cryogenic interferometer of ET and other detectors \cite{penn2003mechanical,hirose2014update,nawrodt2008high}. 
}  \par
The Cosmic Explorer proposal uses silicon ribbons to suspend their test masses and research is ongoing to design the next generation of suspensions \cite{cumming2013silicon}. The properties of the ribbons quoted in the Cosmic Explorer design is assumed when calculating the suspension thermal noise. 
\par
Likewise significant progress is needed to reduce both mechanical and optical losses of the test mass coatings to reach Cosmic Explorer's design sensitivity. 
The progress of these development will be at the forefront of the efforts of next generation of gravitational wave detectors and their results will heavily influence the limits of a lunar interferometer. 

\subsection{Quantum Noise}
Quantum noise is the predominant noise source at high frequencies for Earth bound interferomtric gravitational wave detectors. While frequency independent squeezing is injected in current detectors  \cite{Lough,Acernese2019a,Tse2019} to reduce the shot noise, there are plans for a broadband noise reduction by using filter cavities \cite{PhysRevLett.124.171101} or EPR entanglement \cite{ma2017proposal}. The proposed LION design uses the same optical configuration and laser power (250\,W) as Cosmic Explorer including the injection of 15\,dB squeezing and the application of filter cavities. We have not changed the corresponding parameters because the given design provides a promising noise performance already. A reduction of the radiation-pressure noise at low-frequencies at the cost of high-frequency performance is possible though through the adjustment of laser power. The detector will be quantum noise limited above 2 Hz. \par

\subsection{Seismic Shocks and Asteroid Impacts}
 {Another cause of concern is asteroid impacts on the Moon.
 A large impact near the equipment could potentially lead both to saturated isolation equipment leading to damage to the controls as well as large mirror motion and loss of science mode.
 Reacquiring science mode is a difficult and time consuming process \cite{Staley_2014,Mullavey:12}, 
 However, a similar problem is encountered on Earth with earthquakes. 
 This problem is tackled at aLIGO with a combination of a specialised early warning system for incoming high activity seismic waves and alternative control schemes for high activity times \cite{Biscans_2018}.
 A similar strategy could be copied for any lunar setup.
 Therefore, only a direct hit on the experiment would present a problem. 
 }
 \par
 Based on the data of the still ongoing lunar reconnaissance orbiter, approximately 47\,000 new impact sites had been detected in the area studied (6$\%$ of the lunar surface) in a 1241 day period~\cite{beckman2007mission}. This leads to an estimate of approximately 5 impacts within the detector area (a triangle with 40\,km long sides) in a year. It is foreseen that this will prove to be a regular nuisance, but should still enable long stretches of observation. It is also unlikely that the optics will be hit directly as they will take up a small fractional area of the detector, leading to an estimate of direct hits once in a thousand years. Substance smaller impacts will be significantly under-counted by their very nature, and so it is likely a asteroid shielding solution of the optics will be needed.
 
\subsection{Additional challenges}
 LION will face problems unique to its environment.
 Since the arms are not covered in tubes, Moon dust will eventual stick to the mirrors. 
 Investigations in cleaning strategies on the Moon and the mars are likely to deliver the technologies to reduce this problem, as it is essential for many of the desired experiments and continued exploitation of these bodies \cite{mirror_clean_moon,mirror_clean_mars}.
 \par 
 Furthermore, a mission of this style will require several large payloads to be transported to the Moon and positioned carefully, requiring assembly either by hand or robot.  The feasibility of these strongly depends directly on the lunar missions in the coming decade. However, technology is reaching the stage where commercial missions to the Moon are becoming viable, a promising sign of things to come \cite{chavers2016nasa}.
 \par
 The total mass of the suspended optics is 4\,000\,kg, and the weight of a preisolation stage used at LIGO is 2\,000\,kg \cite{matichard2015advanced}.
 This would leave over 10\,000\,kg as remaining load for both the structure and auxiliary optics, if it was to be carried by the highest capacity rocket today. This would require at least 3 launches for total transport.
 \par
 Gravitational wave detectors produce an high amount of data due to the number of sensors monitoring the numerous degrees of freedom.
 Furthermore, many systems will have to be operated from ground stations.
 A high bandwidth connection will be required to the site.
 Currently a record of 622\,Megabits/s data transfer rate has been achieved by the Lunar Laser Communication Demonstration in orbit of the Moon \cite{boroson2014overview}. 
 With an encoding of 64bits per data point, we would be able to read out about 100 channels at a 100\,kHz sample rate.
 While this may not encompass all of the data producing instruments, it should be sufficient for instrument readout of the main channel and diagnostics of auxiliary channels of interest to above the upper limits of LION's sensitive band.
 \par
 With no need for a vacuum pump, and only small amounts of power needed for alignment controls and isolation, the main power consumption concerns are the telemetry, the high laser power, beam preparation and readout electronics, and the auxiliary sensors. In space missions the power supply is usually covered by a combination of advanced solar panels that reach an efficiency of 30\% and Li-Ion batteries.
  While at the equatorial and mid-latitude regions of the Moon the long lunar nights set limitations for solar power supply, we take advantage of locating the interferometer in polar regions. As shown by \cite{poweratpoles} we receive up to 93\,\% of full or partial sunlight when placing the solar panels on mountains, for example the Malapert mountain, close to the chosen site.
 For 1.4\,kW/m$^2$ of power provided by the sun and a pessimistic factor of 0.5 for the partial sunlight, we gain 0.2\,kW/m$^2$ of solar power, making the power of the system very manageable with a solar panel array of a few tens of square meters.
 \par

\section{Noise Budget}
\label{sec:NoiseBug}
\begin{figure}
 {\includegraphics[width=.6\textwidth,center]{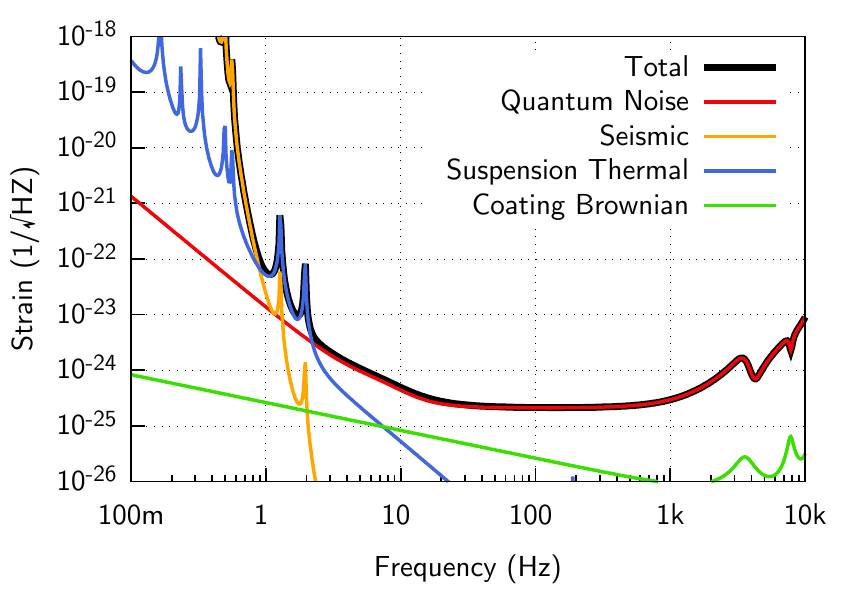}}
\caption{Noise budget for a gravitational wave detector operated on the Moon with an arm length of 40\,km and optics with increased masses at a temperature of 70 K. The parameters used to create this are listed in Table \ref{tab:keyfigs}. The detector is quantum noise limited above 2\,Hz and limited by thermal noise below this until it hits the seismic cliff at 0.8\,Hz. The bandwidth of the detector is therefore approximately from 0.7\,Hz to the kHz. Simulated with gwinc \cite{gwinc}. }
\label{noisebudgey}
\end{figure}
\begin{figure}
 {\includegraphics[width=.6\textwidth,center]{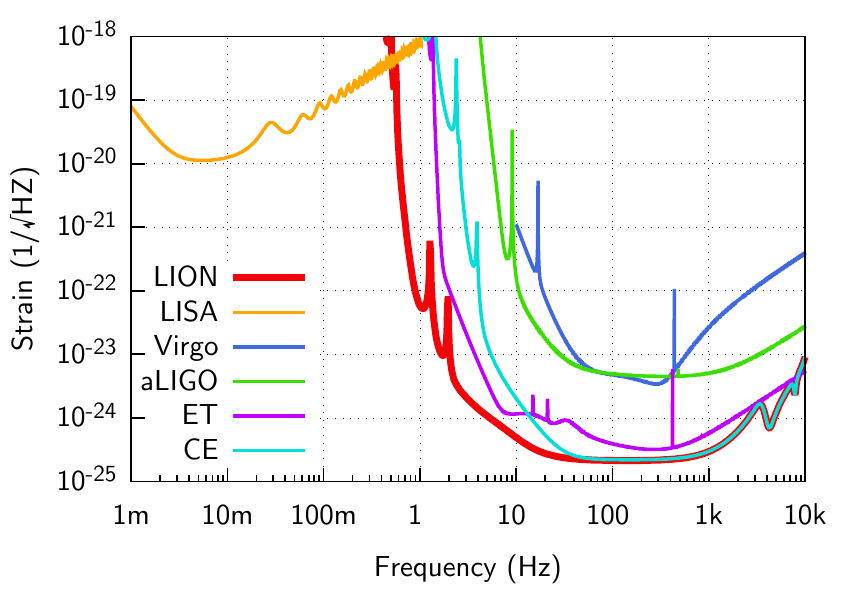}}
\caption{Comparison of the design sensitive of current and future gravitational wave detectors. Our proposed detector, LION, could extend the range of ground based detectors reducing the gap of measurable frequencies between Earth based and space-borne missions .
LION has the highest sensitivity in the detection band from below 30\,Hz and is the only detector able to with sufficient sensitivity for detection between 0.7\,Hz and 2\,Hz. }
\label{noisebudgeycomp}
\end{figure}

{We used the Gravitational Wave Interferometer Noise Calculator (gwinc) \cite{gwinc} software to estimate the noise budget of our proposed LION design shown in Figure \ref{noisebudgey}. Cosmic Explorer (CE2 model, October 2020) was used as a base model, but with the changes discussed in the previous section and listed in Table~\ref{tab:keyfigs} to match the design presented in this paper. The model is currently limited at low frequencies by the properties of the suspension system. Figure \ref{noisebudgeycomp} shows the LION noise budget estimate and compares it with the performance of other current and proposed ground based detectors and the space based detector LISA. Although our estimate is based upon current technology and we possibly overestimate the Moon's seismic activity, LION performs better than ET and Cosmic Explorer at frequencies below 30\,Hz with detection capabilities down to 0.7\,Hz. LION also manages to maintain performance comparable to Cosmic Explorer into the kHz regime.}

\section{Science Case}
\label{sec:Science}
The LION sensitive band, from Figure \ref{noisebudgeycomp},
ranges between 0.7\,Hz and several kHz with peak sensitivity at 10\,Hz.
Several investigations of sources in this frequency range have been conducted
\cite{ETsourcesGair}, showing the great interest in this band, a region where the
current ground based network is blind to.  In general, detecting in a lower
frequency range opens the possibility for observing more massive black holes
and thus allows to study gravity in the strong field regime
\cite{Amaro-Seoane2018}. This is a key element in the testing of general
relativity in more extreme limits. Many new potential sources are predicted to exist in this band, for example, Type 1A
supernovae, which have a gravitational wave frequency signature around 1\,Hz. As discussed in
\cite{Mandel_2018}, their observation may resolve the open question of their
origins.
\par
What makes LION unique, however, is its ability to cover a wide regime of frequencies and characteristic strains of merging binaries which cannot be explored with any other observatory. This forms the focus
for this section. 

\subsection{An Approximation}

\begin{figure*}
          {\includegraphics[width=1.0\textwidth,center]{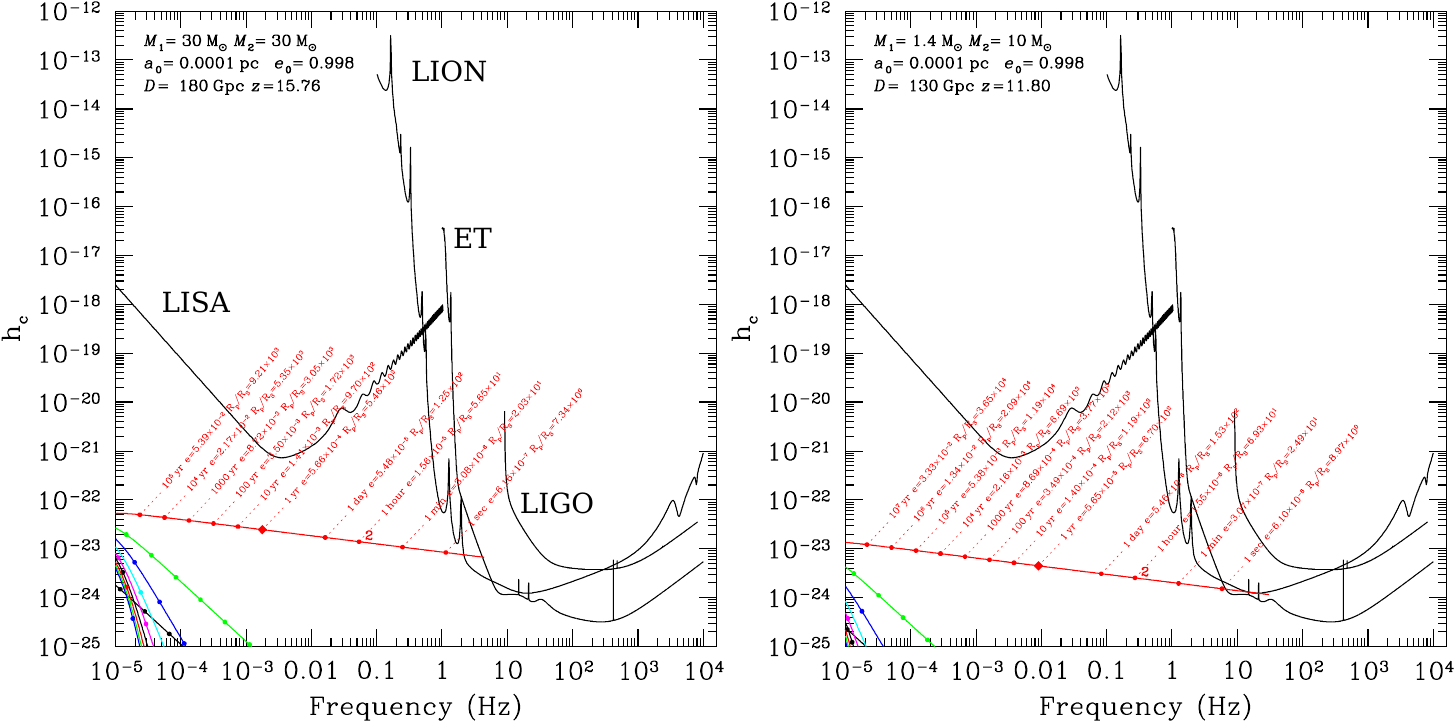}}
\caption
   {
Evolution in frequency of the first harmonics (displayed in different colours)
of two compact binaries in the approximation of Keplerian ellipses of
\cite{Peters64}.  We depict the characteristic, dimensionless amplitude as a
function of the frequency in Hz. The initial dynamical parameters corresponding
to the each source is summarised at the top, left corner of each panel (masses,
initial semi-major axis and eccentricity and distance to the source $D$). From
the left (lower frequencies) to the right (higher frequencies), we show the
sensitivity curve of LISA \citep{Amaro-SeoaneEtAl2017}, then LION, starting at
about 1 Hz, the Einstein Telescope \citep{2012CQGra..29l4013S}, and then LIGO. 
The upper, red harmonic corresponds to $n=2$ in Equation \ref{eq.hn}, which means
that the source has circularised. We also show different moments in the evolution
of the harmonics, and add the information relative to the source at that moment;
in particular the remaining time until merger, the eccentricity and the periapsis
distance $R_{\rm p}$ in Schwarzschild radii $R_{\rm S}$.
The left panel shows a binary of masses $M_1=M_2=30\,M_{\odot}$ at a distance
of 180 Gpc, which corresponds to a redshift of $z=15.76$. This is orders of magnitude
 distance further away than what LISA can hope to observe. The right panel shows the inspiral of a compact
object of mass $1.4\,M_{\odot}$ (e.g. a neutron star) onto a stellar-mass black hole
of mass $10\,M_{\odot}$ at a distance of 130 Gpc ($z=11.80$). Even at this cosmological
distance, the source spends one second in the LION domain. We 
note that no other detector is in the position of resolving these sources.
   }
\label{fig.FarSources}
\end{figure*}

We first give a few examples approximating the evolution in phase-space of the
harmonics of the characteristic strain with the method of Peters et al.\cite{PM63}. This has
the advantage of allowing us to explore the evolution of the eccentricity,
which is rather limited in more realistic waveforms based on computations of
full general relativity.  Once we have an idea of the evolution in phase-space
of these systems, we will investigate in more detail the evolution of the full
waveform and the computation of the signal-to-noise ratio following the method
of Kaiser et al.\cite{KaiserMcWilliams2020}, which we will describe in some detail.
The scheme of Peters et al.\cite{PM63} is based on an approximation of Keplerian ellipses.
If we assume that the orbital parameters change slowly due to the emission of
gravitational radiation, and that the gravitational waves are emitted at the
integer multiples of the orbital frequency, at a distance $D$, the strain
amplitude in the n-th harmonic can be described as 

\begin{align}
    h_n &= g(n,e) \frac{G^2\,M_{\rm 1} M_{\rm 2}}{D\,a\,c^4} \\
        \nonumber &\simeq  1.6\times 10^{-22} g(n,e)
    \left(\frac{D}{1\,\mathrm{Gpc}}\right)^{-1}
    \left(\frac{a}{10^{-2}\,\mathrm{pc}}\right)^{-1} \nonumber \\
    & \left(\frac{M_{\rm 1}}{4\times 10^4\,M_{\odot}}\right)
    \left(\frac{M_{\rm 2}}{10\,M_{\odot}}\right),
\label{eq.hn}
\end{align}

\noindent 
where we have normalised to parameters corresponding to an intermediate-mass
ratio inspiral \citep[IMRI, see
e.g.][]{Amaro-Seoane2020,Amaro-SeoaneLRR,Amaro-Seoane2018,KonstantinidisEtAl2013,Amaro-SeoaneEtAl07},
$g(n,\,e)$ is a function of the harmonic number $n$ and the eccentricity of the
source $e$. In  Figure \ref{fig.FarSources} we give an example of an IMRI and
the inspiral of a binary of two stellar-mass black holes with masses
$30\,M_{\odot}$.

\begin{figure*}
          {\includegraphics[width=1.0\textwidth,center]{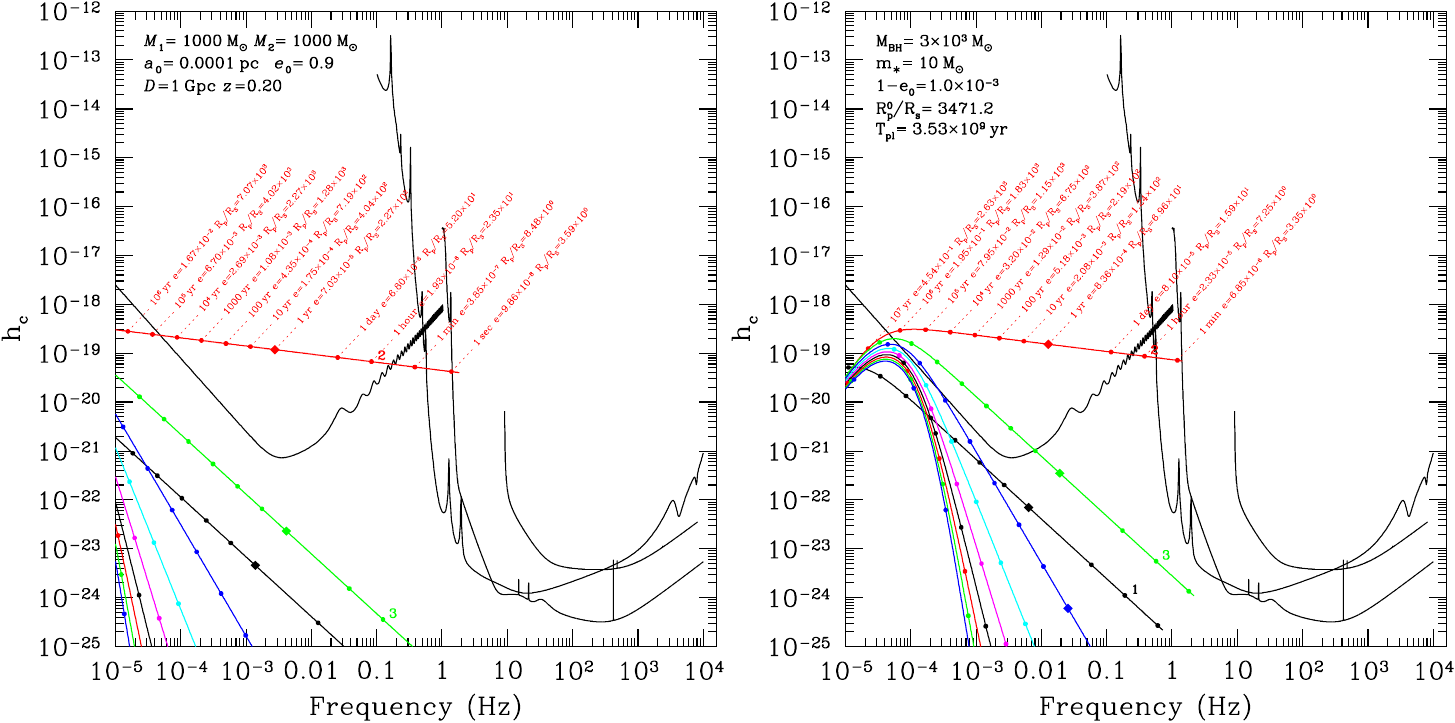}}
\caption
   {
Same as in Figure \ref{fig.FarSources} but for a binary of two
intermediate-mass black holes (left panel) and an IMRI composed of an
intermediate-mass black hole of mass $M_{\rm BH}=3000\,M_{\odot}$ and a
stellar-mass one of $m_{\star}=10\,M_{\odot}$ (right panel). In the latter we
also include information about the initial periapsis distance $R_{\rm p}^{0}$,
as well as the initial time for merger $T_{\rm pl}$.
   }
\label{fig.IMBHs_IMRI}
\end{figure*}

Another interesting example is shown in Figure \ref{fig.IMBHs_IMRI}. A system
of two intermediate-mass black holes (IMBH) and an IMRI which, again, are only
observable by LION at these distances and parameters. As we can see in the left
panel, LION combined with LISA would allow us to do multi-bandwidth
gravitational wave astronomy \citep[see e.g.][]{Amaro-SeoaneSantamaria10},
since a we can observe a source with LISA in the inspiral phase years before
the final merger with a precission of at least seconds, which happens in the
LION domain. This early warning would likely help to enhance the detection of
the merger and ringdown with LION. Thanks to this, we would be in the position
of breaking various degeneracies in the parameter extraction of the source.  As
before, we note that only LION is in this unique position.

As noted in \cite{ChenAmaro-Seoane2017}, there is a need for a deci-Hertz
gravitational wave observatory such as LION to understand the formation of
compact binaries. This is something that LISA and LIGO/Virgo are not in the
position of doing, as explained in \cite{ChenAmaro-Seoane2017}.

\subsection{Realistic waveforms and one comparison to LIGO}

\begin{figure}
          {\includegraphics[width=0.9\textwidth,center]{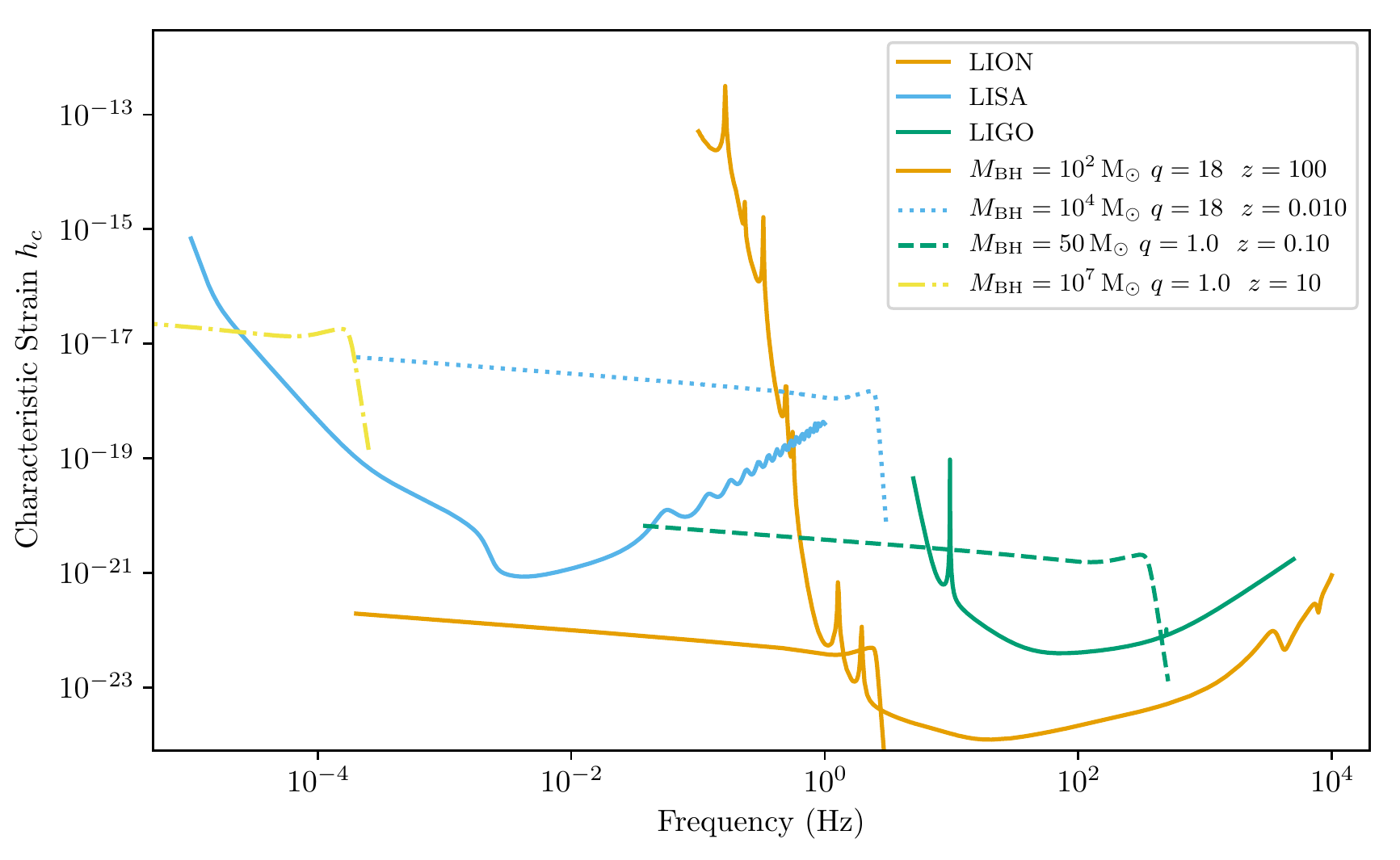}}
\caption
   {
Evolution of four different signals with masses ranging between
$10^7\,M_{\odot}$ and $50\,M_{\odot}$, different mass ratios $q$ (in the legend
we show the mass of the largest black hole) and the assumed redshift. We show
three different observatories (LISA, LION and LIGO). The yellow, dash-dotted
waveform corresponds to a system of $10^7\,M_{\odot}$ at a redsift $z=10$ with
$q=1$; the orange, solid waveform to a binary of $q=100\,M_{\odot}$, $z=100$
and $q=18$; the blue, dotted one to a system of $10^4\,M_{\odot}$, $z=10^{-2}$,
$q=18$ and the green, dashed curve represents a binary of mass $50\,M_{\odot}$,
$z=0.1$ and $q=1$.
   }
\label{fig.detectors_and_sources}
\end{figure}

After our first exploration of the possible interesting systems for LION to
observe, we now address with more realistic waveforms some examples. In
particular, we employ the method of \cite{HusaEtAl2016,2016PhRvD..93d4007K},
commonly referred to as IMRPhenomD, via the approach of
\cite{KaiserMcWilliams2020}. The family IMRPhenomD approaches coalescing
binaries by compounding a post-Newtonian approximation for the inspiral with a
fully numerical relativitistic solution for the merger and ringdown. This hybrid
method allows us to quickly compute  different waveforms. We also consider the
approximation of \citep{2018PhRvD..97b4031H}, ENIGMA, which is a time domain,
inspiral-merger-ringdown waveform model which allows
consideration of moderately eccentric orbits, via the implementation of
\cite{PyCBC2020}. In particular, we display in Figure
\ref{fig.detectors_and_sources} four sources, with one of them only in the LISA
regime as a reference point. 

Probably the most remarkable system is a binary of two black holes of mass
ratio $q=18$ with the heaviest one having a mass of $100\,M_{\odot}$ at a
redshift of $z=100$ (lowermost signal). This binary is only detectable by LION
and shows the relevance of the detector. 

\begin{figure}
          {\includegraphics[width=0.5\textwidth,center]{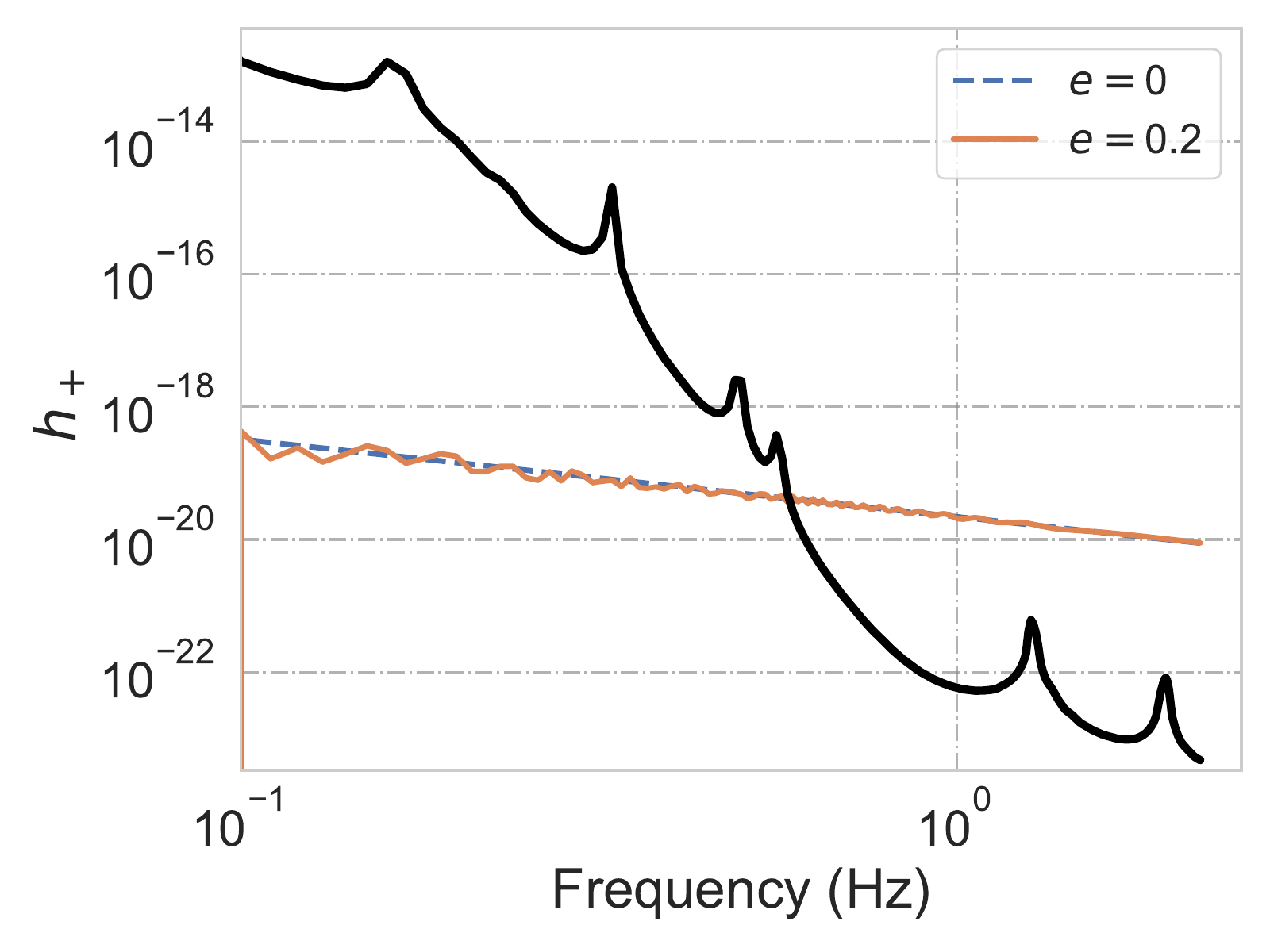}}
\caption
   {
Evolution of the plus polarization $h_{+}$ for a binary of two IMBHs of masses
$M_1=M_2 = 10^3\,M_{\odot}$ in the LION window (solid, black line). We consider
two different eccentricities, one zero and the other $e=0.2$ at the beginning
of the evolution. As the systems evolve, the loss of angular momentum
circularises the eccentric case, which converges towards the circular one. The
binaries are assumed to be at a distance of $1\,\textrm{Gpc}$. For LION, the
circular- and eccentric cases yield $\textrm{SNR} \sim 1855$ and $1875$,
respectively, while for (advanced) LIGO the values are $\sim 3.7$ for the
circular case, and $3.8$ for the eccentric one.
   }
\label{fig.EccentricCase}
\end{figure}

We show in Figure \ref{fig.EccentricCase} a binary of two IMBHs with equal mass
and different eccentricities, as approximated with ENIGMA
\citep{2018PhRvD..97b4031H}.  The case with eccentricity yields a slightly
higher signal-to-noise ratio (SNR) as calculated in the approximation of
\cite{PyCBC2020} for chirping binaries. 

\begin{figure*}
          {\includegraphics[width=1\textwidth,center]{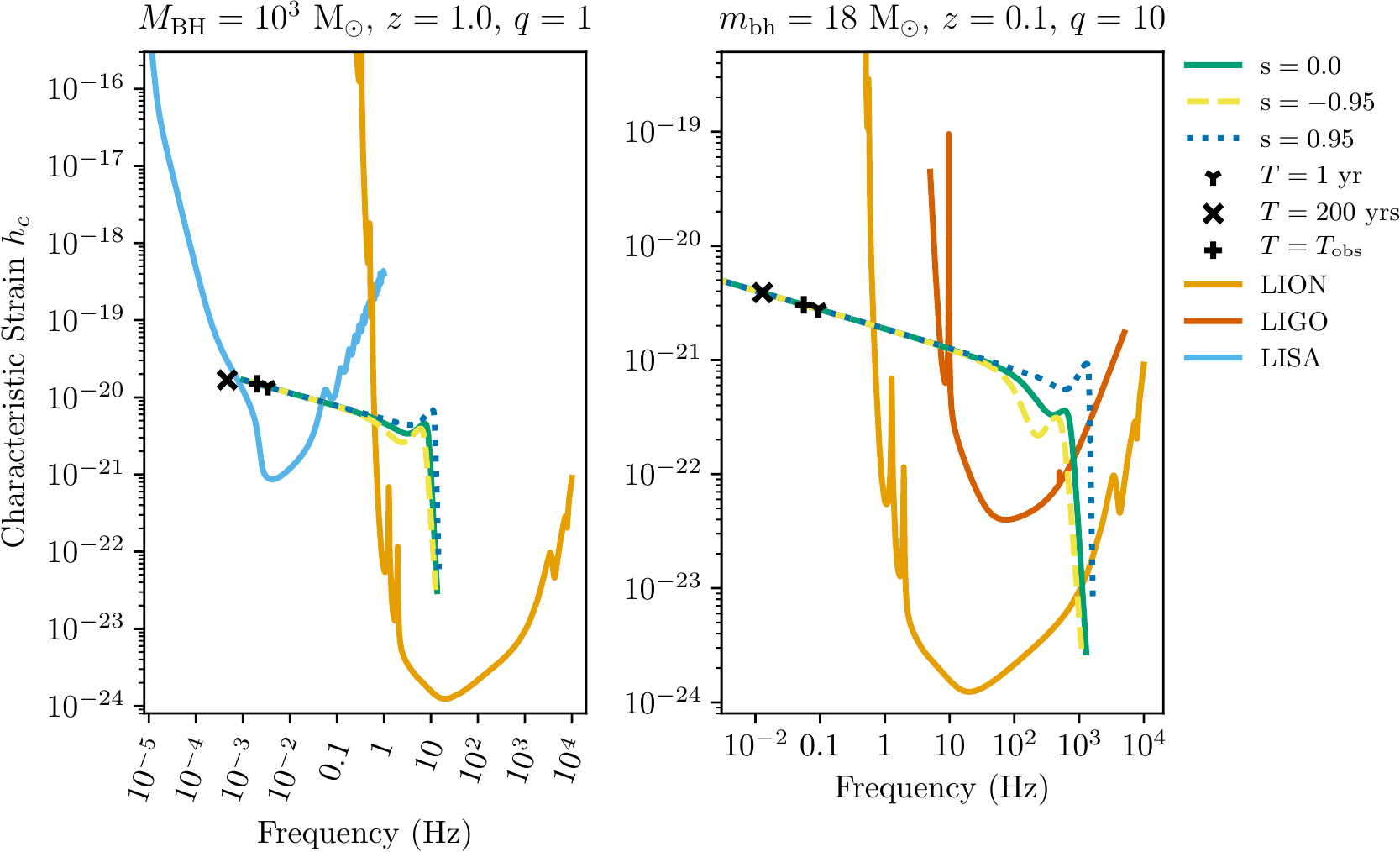}}
\caption
   {
\textit{Left panel: }A binary of two IMBHs of masses $10^3\,M_{\odot}$ at a
redshift $z=1$ for different spin values (see legend on the right) at three
different moments, namely 200\,years before the coalescence, at the
detector's (typical) observation time $T_{\rm obs}$ and 1 year before
the coalescence. We display the sensitivity curve of LISA (blue, left one) and
LION (orange, right one). \textit{Right panel: }Evolution of the waveform for a
binary of two black holes with $q=18$ and the mass of the heaviest one $m_{\rm
bh}=18\,M_{\odot}$, at a redshift $z=0.1$. As in the left panel, we pinpoint
the same moments in the evolution, different spin configurations, the
sensitivity curve of LION (orange, left one) and LIGO (red, right one).
   }
\label{fig.hc_comparison}
\end{figure*}

Because IMRPhenomD features mass ratios up to 18 including spins,
in Figure \ref{fig.hc_comparison} we show the role of the spin $s$ and $q$
for two different systems, a binary of two IMBHs and a lighter one, with
the mass of the heaviest black hole $18\,M_{\odot}$, and $q=10$. As expected,
a mass ratio and positive spin value increases the characteristic strain 
as we approach the end of the evolution, which has an impact on the calculation
of the signal-to-noise ratio.

\subsection{Signal-to-noise ratio contour maps}

To estimate the signal-to-noise ratio for different configurations and
systems, we adopt the approach of \cite{Maggiore2008} and assume that the noise
is stationary, Gaussian and uncorrelated with the signal. If we envisage the
waveforms as vectors in a Hilbert space \cite{Helstrom68}, we can define a
noise-weighted inner product as follows

\begin{equation}
 \left<h'\left|h\right.\right> :=
2\int_{0}^{\infty}{df}\,\frac{ \tilde{h'}(f)\tilde{h}(f)^{*} +
\tilde{h'}(f)^{*}\tilde{h}(f)}{S_{n}(f)}.
\label{eq.}
\end{equation}

\noindent
In the last equation we have introduced $\tilde{h}(f)$ as the Fourier transform
of the time-domain waveform. Moreover, $S_{n}(f)$ is the one-sided noise power
spectral density of the detector, following
\cite{Thorne87,Finn92}, and see Section 2.3 of
\cite{KaiserMcWilliams2020} for a summary and implementation.  If we adopt the
above-mentioned assumptions, the optimal signal-to-noise ratio (SNR) when
filtering the data against $h$ is given by

\begin{equation}
\rho = \sqrt{\langle h_+ | h_+ \rangle + \langle h_\times | h_\times \rangle}.
\end{equation}

\begin{figure*}
          {\includegraphics[width=0.6\textwidth,center]{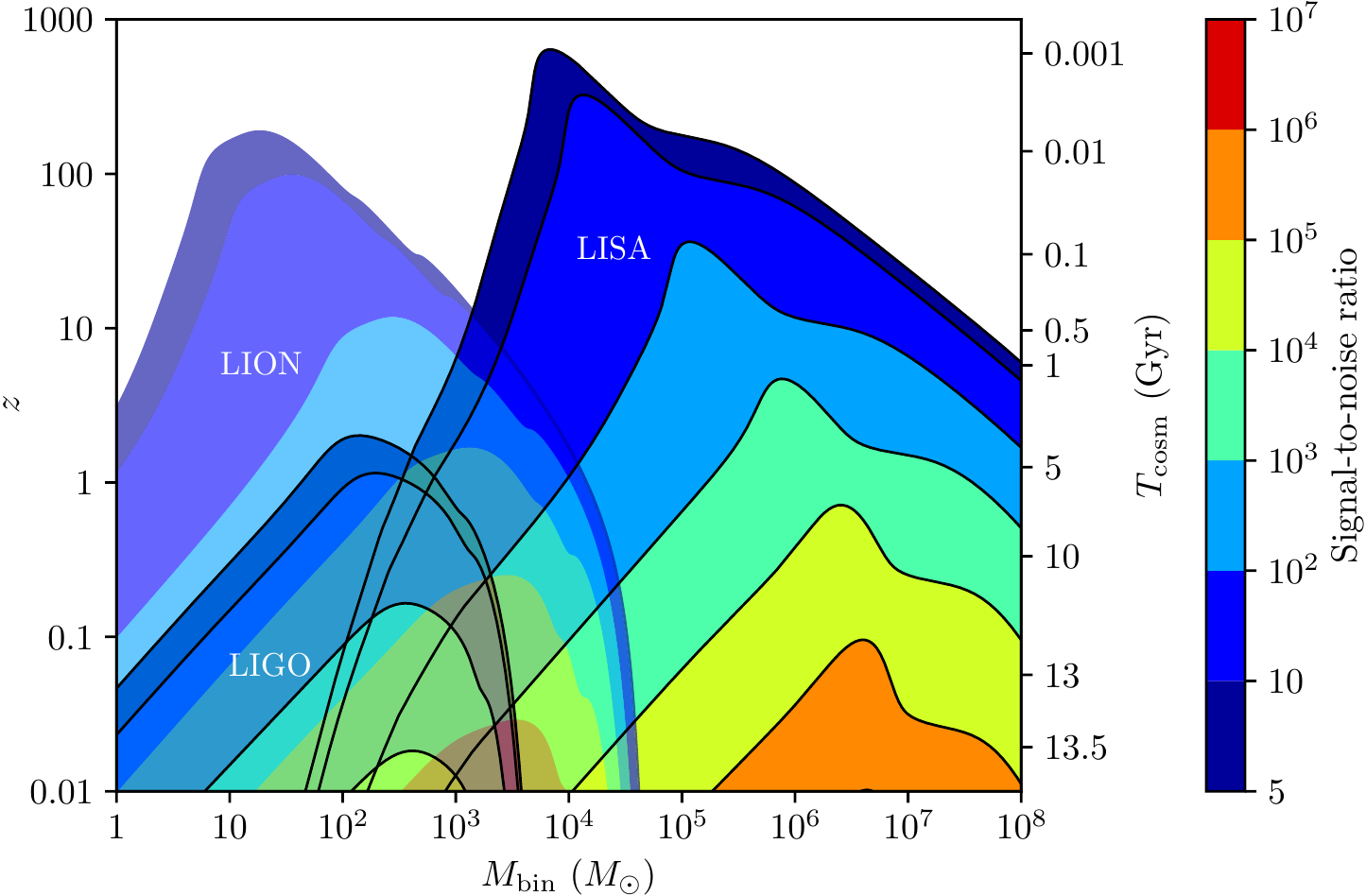}}
\caption
   {
Contours of signal-to-noise ratio as a function of the total mass of the binary system $M_{\rm bin}$
in solar masses and redshift (and cosmological time, in Gyrs). We remark how LION covers a region of
phase-space which is totally inaccessible to both LISA and LIGO, as shown with particular examples in
Figure \ref{fig.detectors_and_sources}.  
   }
\label{fig.Waterfalls}
\end{figure*}

In Figure \ref{fig.Waterfalls} we depict the SNR $\rho$ as a function of the total
mass of the binary $M_{\rm bin}$ and redshift for LION, but also LIGO and LISA for
a reference point. We note that LION is in the position of extracting information about
binaries with masses in the regime necessary to understand the formation and cosmical
evolution of supermassive black holes \citep[see, for a review][]{ColpiDotti2011},
with SNRs values at redshift exceeding 100. 

\begin{figure*}
          {\includegraphics[width=1\textwidth,center]{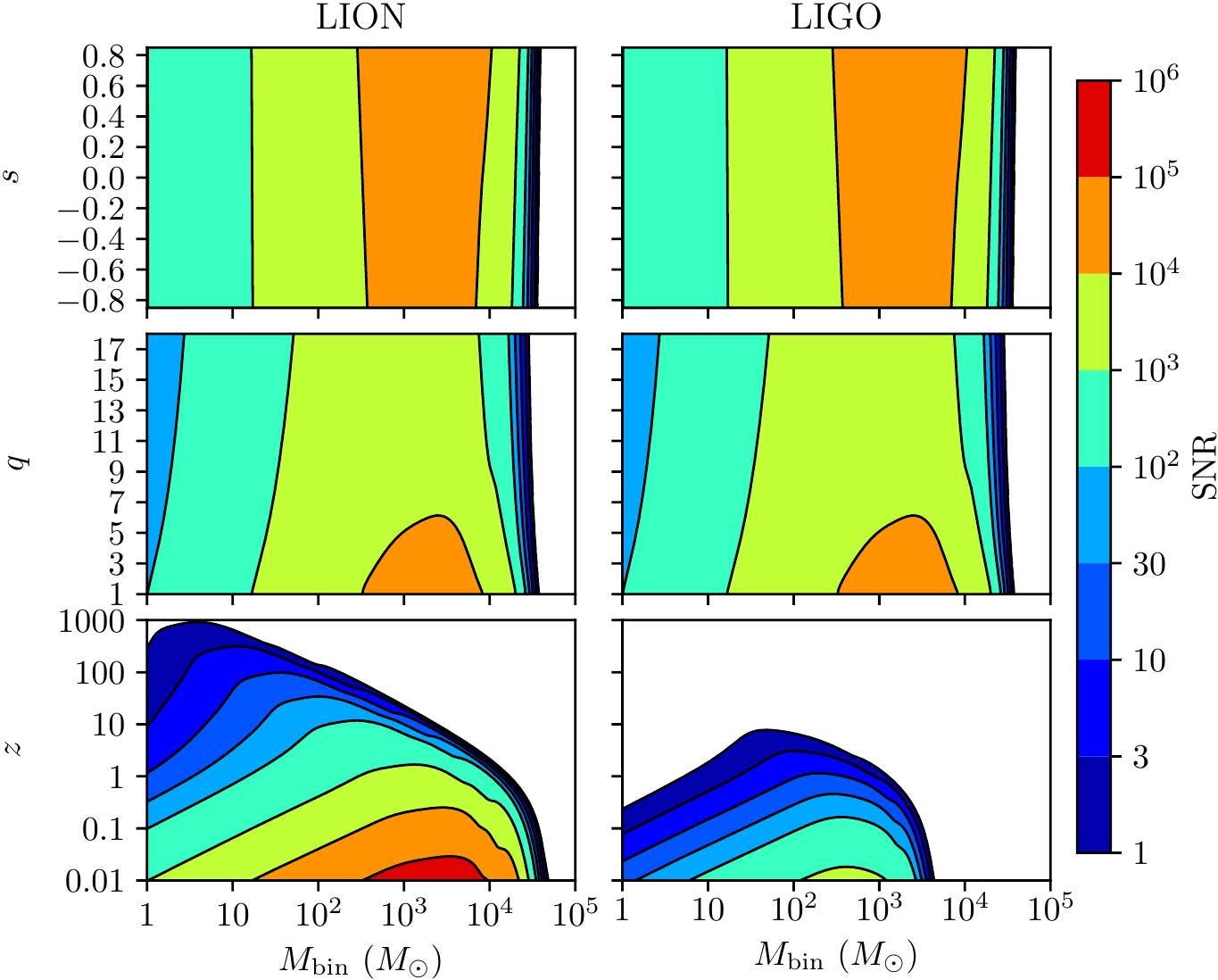}}
\caption
   {
Mosaique of SNR contour maps to show its dependency with the redshift (bottom
panels), as in Figure \ref{fig.Waterfalls}), the mass ratio $q$ ranging between
the values that IMRPhenomD allows for (mid panels, see text) and the spin (top
panels).  We show LION on the left column and LIGO on the right one.
   }
\label{fig.mosaique}
\end{figure*}

Finally, in Figure \ref{fig.mosaique} we show how the SNR depends on the
redshift, mass ratio and spin values for LION and compare it with LIGO as a
reference. We note that, contrary to Figure \ref{fig.Waterfalls}, the contour
map covers SNRs including values from 1.

\subsection{Angular Resolution}

We note that the frequency range of LION allows us to improve the angular resolution of source positioning in the sky, by adding a lunar detector to the LVK collaboration.
The angular resolution $ \Delta \Omega$ is proportional to the increase of the network area A, $\Delta \Omega   \propto 1/A$
\cite{wen2010geometrical}. This means that an average Earth-Moon distance of 380\,000km \cite{Murphy_2013} increases the area A and improves the angular resolution by a factor of 60 compared to a purely Earth based network with maximum radius of 6000km.
\par
This will not only help to enhance the sky mapping of these objects, but also opens a huge possibility to multi messenger astronomy, as we have seen, in particular in Fig.~(\ref{fig.Waterfalls}).
The first counterpart observation GW170817~\cite{PhysRevLett.119.161101}, showed, how much knowledge is gained when observing the emitted gravitational waves and electromagnetic radiation from an event. 
With the combination of improved sky location and knowledge about the time of the merger beforehand, the electromagnetic telescopes can be pointed to the sky location in time to ensure a simultaneous observation. This will significantly increase the upcoming Gravitational-wave Optical Transit Observer (GOTO) array's useful observations \cite{dyer2020gravitationalwave}. 
\par
\section{Cost and Timeline}
\label{sec:costandTime}
The costs for LION can be split into the sections of space operation, including the driving costs of launchers and landers, the communication with Earth, thus the ground segment, and the scientific resources.
Three launches with the commercial Falcon Heavy by SpaceX currently cost 270M Euros \cite{2020falcon}, with additional costs for the landers themselves. 
Space qualification of equipment is often approximately 10 times more expensive than the qualification for ultra high vacuum. Thus, we estimate the infrastructure cost for Cosmic Explorer, without the need of a vacuum system, is 300M Euros \cite{reitze2019cosmic}.
With the additional cost of communication, landers, and automatic installation on the site, as well as all unmentioned human resources, we propose the mission with overall budget in the order of  1,500M Euros.
Future discoveries of new technological requirements, or a significant improvement of the currently changing human spaceflight, however, can influence the cost level in any direction.
We believe that this the right order of magnitude and it is similar to other proposed space based gravitational wave missions,
\cite{Amaro-SeoaneEtAl2017, Danzmann1996}.
\par
With an initial mission proposal by 2023 and a full design study based on new data about the Moon by 2030, simultaneous development of technologies along side Cosmic Explorer's second phase would enable us to begin space qualifying the necessary parts so that all technologies are space qualified by 2048.
This would give time to build the missions in time for 3 launches in the middle of the decade, Assembling would be done robotically and it is hoped that this will be completed within a couple of years. 
The mission lifetime is expected to be 10 years, with the possibility of 5 year expansion.
It is expected that in this time components will wear out and limit operation.
However, if there proved to be sufficient interest, LION can continuously be repaired and upgraded, making an advanced LION detector more likely then the need of disposal.

\section{Conclusion}
We find that a detector built on the Moon with technology of the third generation detectors features a detection bandwidth between 0.7\,Hz and 10\,kHz with detection noise hitting a floor of 3$\times 10^{-24}$\,m/$\sqrt{\textrm{Hz}}$ at 10Hz.  
This is a conservative estimate which is limited by the thermal noise and residual seismic activity of the Moon. %
Since the actual seismic activity of the Moon is still uncertain and we only use an upper estimate of activity,
a thorough test campaign is likely to improve the performance prediction. Nevertheless, LION would make an excellent complement to the existing network, with significant overlap of required technology development. 
\par
The unique detection band presents an opportunity to push the capabilities of gravitational wave astronomy and our understanding of physics. 
The ability to measure binaries of seed black holes to a redshifts of 100 and higher provides a unique glance at the cauldron where stars and galaxies were formed and would shed light on the question on how massive black holes assemble their mass and grow over cosmic timescales \citep[see e.g.][for a review]{ColpiDotti2011}. 
We also find that many of the sources present in the LISA band can be picked up in the LION band to observe final merger achieving the long held dream of multi messenger astronomy. Doing so will break many of the degeneracies of the parameter space so that we can better understand their formation channels \citep{ChenAmaro-Seoane2017}. In this regard, it can be envisaged as a true deci-Hertz bridge between space borne observatories and ground based ones.
Moreover, having a much wider detection array at low frequencies is imperative for the coordination between electromagnetic and gravitational observations needed to glimpse the final stages of an inspiraling system with an optical telescope. 
\par
The Moon is perhaps one of the best available environments to operate a gravitational wave detector in our solar system.
While we show in this paper that current or currently designed technology enables the low frequency detector, its installation on the lunar surface requires significant technological development. However, these lie not too far beyond the horizon of humanities progress. 
We believe that, once humanity starts exploiting the Moon for science and resources, a gravitational wave observatory is an excellent candidate for the first large scale infrastructure experiment on the lunar surface.

\par

\section*{Acknowledgements}

This paper builds on a student project from an lecture week by the International Max-Planck Research School for Gravitational Waves (IMPRS-GW) in 2018. 
Therefore we would like to thank the IMPRS-GW for
organisation and Karsten Danzmann for providing the space necessary for the
development of this research. 
We thank Stefan Danilishin and James Lough for
the discussion of the technical and the environmental challenges. We are also
thankful to Christoph Vorndamme for discussion on the initial noise budgets.
We thank Frank Ohme for his comments and suggestions, as well as Andrew Kaiser for comments and for making his approach publicly available.
Finally, we would like to thank Harald L\"uck and Jens Reiche for useful feedback on the
manuscript.  
\par
We also acknowledge that the work of the authors Bischof, Carter, Hartig and Wilken was supported by the IMPRS-GW grant
and the funding by the Deutsche Forschungsgemeinschaft in
the Collaborative Research Center SFB1128 Relativistic Geodesy and
Gravimetry with Quantum Sensors (geo-Q) at Leibniz Universität Hannover.
Bischof, Carter, Hartig and Wilken gratefully acknowledge support by the Deutsches Zentrum für Luft- und Raumfahrt (DLR) with funding from the Bundesministerium für Wirtschaft und Technologie (FKZ 50OQ1801).
Pau Amaro Seoane acknowledges support from the Ram{\'o}n y Cajal Programme of the Ministry
of Economy, Industry and Competitiveness of Spain, as well as the financial
support of Programa Estatal de Generación de Conocimiento (ref.
PGC2018-096663-B-C43) (MCIU/FEDER), and the COST Action GWverse CA16104. This
work was supported by the National Key R\&D Program of China (2016YFA0400702)
and the National Science Foundation of China (11721303).

\section*{Author contributions}

The initial idea of LION was developed by the authors Bischof, Carter, Hartig and Wilken at a lecture week in 2018, where Amaro Seoane was supervising, run by the International Max Planck Research School for Gravitational Waves (IMPRS-GW).
 Bischof, Carter, Hartig and Wilken
followed up the lecture week with a deeper exploration of the required mission parameters, noise curves and technical discussion of the paper. 
The work on sources in the detection band was performed by Pau Amaro Seoane.

\bibliographystyle{vancouver}

\end{document}